# The Harmony Index: a Utilitarian Metric for Measuring Effectiveness in Mixed-Skill Teams

**Darryl Roman**
Learning Sciences Cluster
University of Central Florida
darrylroman@knights.ucf.edu

**Noah Ari**
Learning Sciences Cluster
University of Central Florida
noahari@knights.ucf.edu

**Johnathan Mell**
Learning Sciences Cluster
University of Central Florida
johnathan.mell@ucf.edu

**Abstract**

As teamwork becomes ever-more important in a new age of remote work, it is critical to develop metrics to quantitatively evaluate how effective teams are. This is especially true with mixed-modality teams, such as those that include a human and an agent or human and robot. We propose a novel utilitarian metric, the *Harmony Index*, which quantifies the effectiveness of team members by classifying them into four subtypes based on the result of their teaming on overall effectiveness. This index is evaluated using a real-world dataset of over 1 million interactions, and potential future uses of this index are explored in the realm of team science.

## 1. Introduction

Efficient work benefits from teamwork. As the world has endured the turmoil of remote work, new team members, and increased automation, determining how to build efficient teams has taken on a new urgency. In this paper, we examine fundamental metrics required to build these teams, with broad applicability to teams of virtual agents and humans (as well as human-human teams). Metrics of team effectiveness are critical to further develop methodologies for determining optimal team composition, whether it be human-human, human-agent, or human-robot. Put directly: in order to establish effective teams, there must needs be quantitative metrics to measure effectiveness.

To wit, effective teams will need metrics assessing task proficiency to determine if they are successful and improving. But, more than that, these metrics need to be easily understood, comparable, and derived from sound mathematical principles so that they can be applied to agent/robot models as well as understood by the team-builders who would use them.

While previous research (DeShon et al., 2004) has proposed a multilevel model which calculates individual and team metrics to determine team performance, it has not proposed a true joint measure that easily represents effectiveness. In this work, we propose initializing team organization with a novel utilitarian methodology for organizing virtual and human agents based on a single Harmony Index. While this methodology can be applied to human-agents specifically (and indeed, has been designed with that end in mind), it can apply to teams of any modality. We define this index by analyzing the pairwise benefit of teams (see Section 4) on task accomplishment rates and categorizing each agent pair on four harmonic classes based on degree of benefit to each agent: Harmony, Uplift, Depress, or Discord. We call the result of this analysis the *Harmony Index*.

The Harmony Index is an easy-to-use, straightforward measure of team effectiveness. Our methodology involves paring down the multiple varied and potentially compounding interactions into a single scalar number. We examine our index using a real-world dataset, and we show how the Harmony Index represents a useful way of examining effectiveness in a complex domain. Our analysis is based on proven success rates on the intended task, which is a complex 5-vs.-5 competitive online game. We examine the Harmony Index for dyads on the same team, which provides an optimal utilitarian starting point for each pairing. We also show how this same index can be calculated for and applied to larger teams. The details of this calculation are expounded upon in Section 4.

Having the data for a dyad is of particular interest in designing larger teams. In the case of this dataset, we have the actaul real-world data for the 5-person success rate. However, this is unlikely in many other scenarios. Often, new teams may be composed from smaller subgroups. It is not unreasonable to suppose that dyads may have worked together before, and this information can be used to build effective larger teams. However, by pure stastical improbability, having the data for a 5-person (or larger) team a priori is unlikely. Therefore, having the Harmony Index provides clear benefit from the perspective of team construction.[1]

In the remainder of this paper, we describe the quantitative framework used to derive these Harmony Indices, the resulting 4 classes of agent-agent relationships, and a methodology for subsequent iterations in order to adapt team membership over time or as additional data points are generated.

## 2. Task Description

The dataset used to analyze and generate the Harmony Indices is generated from the Mobile Online Battle Arena (MOBA) game, Heroes of the Storm (HotS). HotS is a competitive team versus team game with various types of cooperative tasks. These tasks can be completed to improve either teams' likelihood of achieving the Success/Fail final objective. The teams compete to complete these objectives before the other team. Each player/agent has unique skills and/or abilities that create a range of codifiable interactions between agents, which can be measured in their intersection. A set of 90 agents generated a range of 4,005 distinct pairwise agent interactions. Given a binary Success/Fail final objective, an effective success rate of 50% defines the base value of a successful agent or agent pairing.

The game uses a system of ranks and match making rating (MMR) to denote strong or successful players and teams. These ranks and ratings are determined by success rate over teams with similar ratings. Teams with higher success rates will be matched against teams with similar ratings until an equilibrium is reached. Therefore, a 50% success rate is necessary to maintain a stable ranking.

In this environment, each agent is also given a specialized set of traits and/or skills which allow it to excel in accomplishing certain tasks and lacking on others. These specializations will not be directly considered in this work, but can be seen as analogues for differing skill sets found in real-world teams. We therefore do not integrate a "skills" measure into our Index, as we are attempting to establish a clean baseline based solely on overall task effectiveness. We leave the influence of these specializations to future work.

All data used in this study was collected voluntarily from players of Heroes of the Storm who installed a local client which uploaded game data to a centralized server.

## 3. Related Work

Team performance has been studied in the space of MOBAs in numerous other studies. For example, Sapienza et al, 2018 closely examined effects to performance of individual players over time and their improvement through training. Kozlowski and Klein, 2000 ask if team effectiveness is just an average of effectiveness or if it arises from an array of specializations at various tasks in the overall objective. Kim et al, 2016 examined the methods through which players construct a team in the interest of completing adjectives in the interest of successfully completing the final objective before the competing team.

The domain in Kim et al, 2016 is a similar MOBA to the one from which our data is extracted. However, they focus primarily on the effects of the player's expertise, self-efficacy, and cohesion and their effect on the team. In this paper, we will be aligning more with Gorman, Grimm, and Dunbar, 2018, who suggest that metrics for team effectiveness are better measured at the level of the full system. As such, we are presenting a methodology for calculating a unified index metric which reduces individual player's effects to determine the neighbor overlap, which Damacharla et al., 2018 defines as the performance effect of one agent upon another agent, of a particular agent pairing.

## 4. Equations

### 4.1. Solo & Duo Probabilities

We use the following definitions in our equations:

- $P(A)$: Probability of a win given that Agent A is on the team we are examining
- $P(B)$: Probability of a win given that Agent B is on the team we are examining
- $P(A \cap B)$: Probability of a win given that Agents A & B are both on the team we are examining
- $P(A \cap \overline{B})$: Probability of a win given that Agent A is a member of the team we are examining, but Agent B is not.
- $P(B \cap \overline{A})$: Probability of a win given that Agent B is on the team we are examining, but Agent A is not
- n: Number of team members whose relationship we are testing
- Target Success Rate > 50%

---
[1] We acknowledge that simply because two dyads are effective, there is no reason to suspect that the four-man team assembled from their composition would necessarily be effective. We discuss this as further research.

## 4.2. Harmony Index

The Harmony Index measures the strength of a set of team-members, so in order to get a single scalar value we constructed a general formula that expresses this. First we find the percent increase in likelihood of a Win (or success state) given that an agent or player is on a team with each other relevant or examined agent when compared to the odds of each agent winning without those team members. Then we repeat this for each examined team member and take the geometric mean where n is the number of examined team members.

$$\sqrt[n]{\prod_{y \epsilon T} \prod_{x \epsilon T} \frac{P(x \cap y)}{P(x \cap \overline{y})}}$$

Equation 1: Generalized Harmony Index Calculation for a team T of size n

$$\sqrt{\frac{P(A \cap B)}{P(A \cap \overline{B})} * \frac{P(A \cap B)}{P(B \cap \overline{A})}}$$

Equation 2: Harmony Index Calculation for a team of size 2

The generalized formula (shown in Equation 1) in the case of two-member teammates reduced to the formula shown in Equation 2, which gave us a very strong indicator of how each agent was either uplifted or depressed by each possible agent as a pairing. $P(A \cap B)$ is our clean baseline for its individual effectiveness with all other agents by excluding all interactions that intersect with Agent B. This allows us to directly examine how much agent A's rate of success is impacted by B's presence on the team by calculating $P(A \cap \overline{B})$ which presents us with a score indicating strength of the relationship from the perspective of agent A. We then compute the same value for Agent B with the same paired effectiveness over $P(B \cap \overline{A})$ providing its own effectiveness when not paired with Agent A. This provides a view of the strength of the relationship from the perspective of Agent B.

Once the values are determined for the agent pairings, these values are then combined via a geometric mean in the Generalized Harmony Index Calculation to produce a value for the strength of relationship from the perspective of the team as a whole.

## 5. Relationship Classes

Based on theoretical patterns of interaction and the formulae demonstrated in section 4, we define a resultant 4 classes of agent-agent relationship. Each class represents the task success rate of each half of the relationship and, in the case of imbalance, the effect on the resultant pairing's effectiveness rate. These classes are indicated by combinations of Wins (W), where an agent improves its performance, or Losses (L), where the agent lowers its performance.

### 5.1. Harmony (WW) & Discord (LL)

The first 2 classes represent an interaction where both agents receive the same effect of their collaboration. In the case of a Harmonious pairing (WW) both agents show an increase respective to their individual task success rate and, in the vast majority of cases, these Harmonious pairs found a Success Rate above the target success rate of > 50%. In our data set, the highest Harmony Index was between agent 40, P(A) = 46.44%, and agent 66, P(B) = 48.67%, who had a Harmony Index of 1.28 and a joint task success rate of 60.81%. This represents an increase in their average win-rate of 13.26%.

We defined a Discordant pairing (LL) as a pairing where the joint task success rate is lower than the respective success rate of each individual agent. This class shows that these pairings are not effective at improving their effectiveness and, in fact, these agents seem to interfere with each other. In our data set, the lowest Harmony Index was between agent 53, P(A) = 41.55%, and agent 71, P(B) = 42.70%, with a Harmony Index of 0.72 and paired success rate of 30.49%. This represents a decrease in their average win-rate of 11.63%.

### 5.2. Uplift (WL+) & Depress (WL-)

The Uplift class demonstrates a more subtle benefit to the pair. In this case the success rate of one of the agents increases while the other decreases. In the dataset, the highest Harmony Index that confirmed to this pattern was 1.11: there is an increase to the pairwise task effectiveness rate. For example, in the pairing of agents 7 and 39, the pairwise success rate is 57.13%. But, P(A) = 46.28% and P(B) = 58.61%. So. while agent 39 is losing some of its effectiveness, 0.97%, in the pairing, it is also uplifting agent 7 by a dramatic 23.44%, causing overall mutual benefit.

On the other hand, the Depress class represents a possibly economically divisive set of pairings. In these cases the success rate of the agents is negatively impacted by the pairing, but one of the agents is still achieving an increase to their effectiveness. In the pairing of agents 53 and agent 37 which has a P(A∩B) = 42.56%. Agent 53, P(A) = 41.53, achieves a 2.46% increase in its task effectiveness while agent 37, P(B) = 52.94, suffers a 19.61% decrease compared to its rate without agent 53. This creates a Harmony Index of 0.90, well below the neutral index of 1. To use layman's terms, we're "spending dollars to buy cents". While that is generally not an effective solution, there will likely be scenarios where task specializations or multiple vectors for success are involved where this could still be an effective trade-off.

### 5.3. Class Distribution

Our dataset consisted of 1.67 million data points across 90 agents. This data was collected over a period of 30 days. We did not include any agent pairings with fewer than 1,000 intersecting data points, in order to ensure sufficient statistical power in our results.

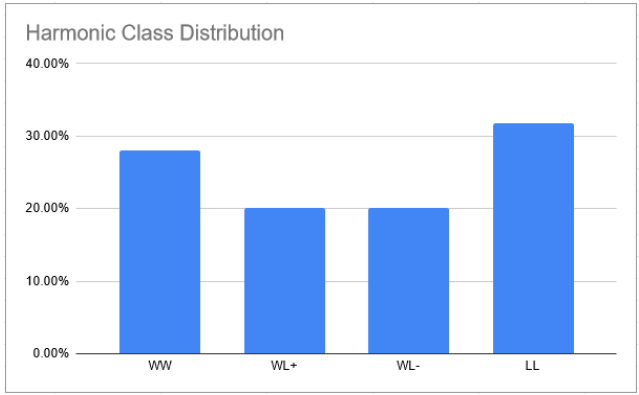

Figure 1: Harmony Index, Class Distribution

Our cleaned results found 812 purely Harmonious agent interactions, 581 Uplifting, 579 Depressed, and 918 purely Discordant interactions. This distribution is represented in Figure 1.

## 6. Results & Discussion

### 6.1. Results Plots

In order to validate our Harmony Index as an ecologically valid metric, we examined the distribution for normalcy. The distribution follows a normal curve, which can be ascertained from the occurrence plot in Figure 2.

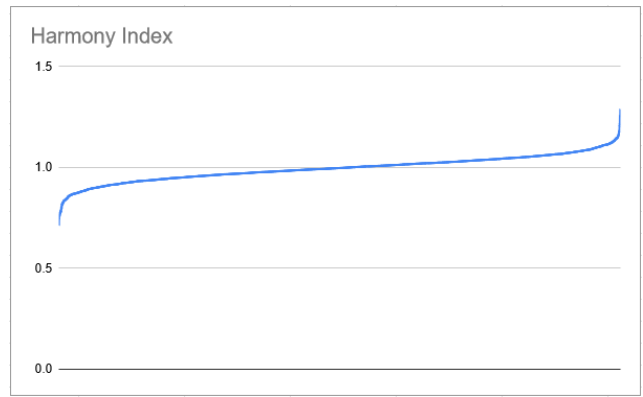

Figure 2: Harmony Index, Values & Distribution

The data conform to an expected distribution with the vast majority of values near a neutral Harmony Index with the 2nd and 3rd quartiles within a range of -.043 to .037, as shown in Figure 3.

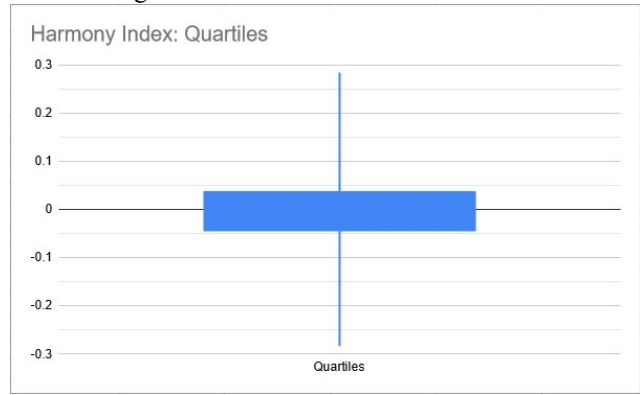

Figure 3: Harmony Index, Quartiles

### 6.2. Notable Edge Cases

There were 17 agent pairings that we felt were notable in that their Harmony Index > 1 and they are categorized as Harmonious pairs (WW), but their chance of winning together fell below the minimum target success rate of 50%. Therefore, though the agents did prove to be beneficial to each other, their combination still failed to be an effective pairing. For example, Agent 7 has a P(A) = 46.29%, Agent 68 has a P(B) = 44.19%, and together they are a Harmonious pair with a Harmony Index of 1.05. However, their P(A∩B) = 47.23% is still below the Target Success Rate of > 50%.

In some sense, this represents a subclass of our Harmonious pairs, and we feel this is particularly interesting because they may be viewed as suboptimal by casual observers, but they still provide clear benefit.

In particular, we are interested in the intuition of real humans interacting with such teams. Prior work has shown that humans do attend to the benefit of team performance over individual performance. However, some studies such as Myers (2007) and Achternbosh et al., 2014. have shown that in some cases human agents will value their own benefit over team benefits, or even depress the team's effectiveness at the cost of their own effectiveness. However, these studies have been on games with less tightly-coupled teams and we hypothesize that our very complex domain might show opposite effects, or more subtle interactions with the aforementioned subclass. We believe that the construction and validation of the Harmony Index is a key first step in explore these HCI topics.

# 7. Conclusion & Future Work

Human, computer, robot, and mixed modality teams have moved out of the realm of science fiction into use in common everyday functions. In order to optimize these team structures we must be able to quantify the results of these structures in a way that will be useful and iterable in order to improve the structures over time. We offer the Harmony Index as a functional building block for the building and optimization of teams by working out the top level relationships of the team components.

However, the Harmony Index is only a single piece of that puzzle and a starting point from which further research can be performed. In order to move forward and improve the results of iterative cycles we must understand the details of the individual components as well as unintentional effects that can arise from using a solely utilitarian perspective on agent teaming.

## 7.1. Abuse of the Depress Class

An example of an unintentional effect is the small subset of edge cases in the Depress class in which one agent of the pair extracts a significant benefit to its own success rate from the pairing, but does so at a significant cost to the success rate of the paired agent causing the previously successful paired agent to fall below the Target Success Rate. We hypothesize that this sort of egoist pairing would be susceptible to abuse, whether it be computer, human, or robot, where one agent is willing and/or able to heavily weight its own welfare over the welfare of its paired agent or team.

It is overly optimistic to believe that all parties in a human, computer, robot team will be utilitarian in nature. The realistic viewpoint is this egoist behavior has been shown in real world scenarios. Uhlmann and Barnes, 2014 demonstrated the occurrence as well as benefit to the individual over the team. Apt and Schäfer, 2014 provides a framework for incorporating philosophical concepts into effectiveness calculations.

Therefore, we suggest further study is warranted on the integration of concepts such as the Prisoner's Dilemma, Nash Equilibrium, Pareto Optimization, and the psychology of intention in order that agents are able to account for egoist behaviors in prospective teams and teammates and whether this behavior can be predicted from partial data points to prevent abuse by egoist agents.

## 7.2. Visualization

As this Harmony Index is a useful method of deriving pairwise relationships we believe that we can apply a forced Network graph visualization to the network of agents in order to generate a single visual representation of the wholesale relationships and their clustering tendencies.

Through this method researchers will be able to map out large datasets of relationship data in a single visual representation that inherently demonstrates patterns and outliers.

## 7.3. Future Opportunities

One final area of interest for further study is the addition of the specializations aspect of the various agents. Each agent in the source game is equipped with skills and abilities that give it specialized skills in various objectives to the game itself. Some have direct impact abilities, some reduce the aspects of opposing agents, some interact directly with smaller objectives, or may give nearby agents an increase in their own abilities. Our Harmony Index does not consider these specializations in its algorithm for neighbor overlap.

Kim et al, 2016, as well as Evertsz and Thangarajah, 2020 proposed that in order to optimize team performance in multi-agent systems the participating human agents must understand what their individual objectives are and how to accomplish them. We believe this is testable as each of our agents has a defined role in the game we should be able to regress the results of the Harmony Index along with the matching roles to determine whether the specialization is a factor in the effective success rate.

# Acknowledgements

The authors thank HotsLogs.com for the use of their dataset.